\documentclass{llncs}
\usepackage{latexsym,color,graphics,times}
\usepackage{diagrams}
\usepackage{url}
\usepackage{epsfig}
\usepackage{amssymb}
\usepackage{amsmath}
\usepackage{amsfonts}

\begingroup
\catcode`\~=11
\gdef\urltilde{\lower 0.6ex\hbox{~}}
\endgroup

\newcommand{\A}{\mathcal{A}} \newcommand{\B}{\mathcal{B}}
\newcommand{\C}{\mathcal{C}}

 \renewcommand{\L}{\mathcal{L}}
\newcommand{\M}{\mathcal{M}}

\title{Data Base Mappings and Theory of Sketches}
\date{}
\author{Zoran Majki\'c}
\institute{ International Society for Research in Science and Technology\\
 PO Box 2464 Tallahassee, FL 32316 - 2464 USA\\
 \email{majk.1234@yahoo.com},\\
~~~~http://zoranmajkic.webs.com/}

\newtheorem{propo}{Proposition}
\begin{document}

% \firstpage{1}

\maketitle
\begin{abstract}
In this paper we will present  the two basic operations for database
schemas used in database
 mapping systems (separation and Data Federation), and we will explain
 why the functorial semantics for database mappings needed a new
 base category instead of usual $\textbf{Set}$ category.\\
 Successively,  it is presented  a definition of the graph $G$ for a schema database
 mapping system, and the definition of its sketch category
 $\textbf{Sch}(G)$. Based on this framework  we presented
 functorial semantics for database mapping systems with the new base
 category $\textbf{DB}$.
      \end{abstract}
% per tabella di contennuto
%\newpage
%\tableofcontents
%\newpage
%
\section{Introduction}
Most work in the data integration/exchange and P2P framework is
based on a logical point of view (particularly for the integrity
constraints, in order to define the right models for  certain
answers) in a 'local' mode (source-to-target database), where a
general 'global' problem of a \emph{composition} of complex partial
mappings that involves a number of databases has not been given the
correct attention. \\
This work is an attempt to give a correct  solution for a general
problem of complex database-mappings and for high level algebra
operators for database schemas (separation, Data Federation),
preserving the traditional common practice logical language for
schema database
mapping definitions.\\
Only a few works considered this general problem
~\cite{MBDH02,AlBe02,DaBK98,MeRB03}. One of them, which uses a
category theory ~\cite{AlBe02}, is too  restrictive: their
institutions can be applied only for  \emph{inclusion} mappings
between databases.\\
There is a lot of work for sketch-based denotational semantics for
 databases \cite{LeSp90,RoWo92,DiCa95,JoRo00}. But all of them
use, as objects of a sketch category, the elements of an ER-scheme
of a database (relations, attributes, etc..) and not the
\emph{whole} database as a single object, which is what we need in a
framework of inter-databases mappings. It was shown in \cite{Disc97}
that if we want to progress to more expressive sketches w.r.t. the
original Ehresmann's sketches for diagrams with limits and
coproducts, by eliminating non-database objects as, for example,
cartesian products of attributes or powerset objects, we need
\emph{more expressive arrows} for sketch categories (diagram
predicates in \cite{Disc97} that are analog to the approach of
Makkai in \cite{Makk94}). Obviously, when we progress to a more
abstract vision where objects are the (whole) databases, following
the approach of Makkai, in the new basic category $\textbf{DB}$ for
databases, where objects are just the database instances (each
object is a set of relations that compose this database instance),
we  obtained much more complex arrows. Such arrows are not simple
functions, as in the case of base $\textbf{Set}$ category, but
complex trees (operads) of view-based mappings. In this way, while
Ehresmann's approach prefers to deal with few a fixed diagram
properties (commutativity, (co)limitness), we enjoy the possibility
of setting full
relational-algebra signature of diagram properties.\\
%The other are not adequate because are to much generic:
%a binary domain relations between databases are to much
%wide to capture the meaningful semantics of a more structured
%mappings between databases, generally defined by using a logical
%implication between corresponding \emph{views} of different
%databases.\\
This work is an attempt to give a correct solution for this problem
while preserving the traditional common practice logical language
for the schema database mapping definitions.\\
The   \emph{instance level} base database category $\textbf{DB}$ has
been introduced first time in Technical report ~\cite{Majk03}, and
used also in
 \cite{Majk03f}. General information about categories the reader can find in classic books ~\cite{McLn71}, while more information about this
 particular database category $\textbf{DB}$,
  with set of its objects $Ob_{DB}$ and set of its morphisms
  $Mor_{DB}$, are recently presented  in \cite{Majk04AOT}.
  In this paper we will only emphasize some of basic properties of
  this $\textbf{DB}$ category, in order to render more selfcontained this
  presentation.\\
  %\begin{enumerate}
Every object  (denoted by $A,B,C$,..) of this category is a
  database instance, composed by a set of n-ary relations  $a_i\in A$, $i= 1,2,...$  called also "elements of
  $A$". \\
  We consider the views as a universal property for databases:
  they are the possible observations of the information contained in
  an  instance-database, and we may use them in order to establish an equivalence relation
  between databases.\\
%\begin{enumerate}
   In  ~\cite{Majk03} has been defined  the power-view operator $T$, with domain and codomain equal to the set of all database
  instances, such that for any object (database) $A$, the object
  $TA$  denotes a database composed by the set of \emph{all views} of $A$. The object $TA$, for a given database instance $A$, corresponds to
   the  quotient-term  algebra $\L_A/_\approx$,
  %of the "select-project-join + union" query   language (SPJRU).
  where carrier is a set of equivalence classes of closed terms of a
well defined formulae of a relational algebra, "constructed" by
$\Sigma_R$-constructors (relational operators in SPJRU algebra:
select, project, join and union) and symbols (attributes of
relations) of a database instance $A$, and constants of
attribute-domains.\\
 Different properties
of the base $\textbf{DB}$ category are considered in a number of
previously
published papers \cite{Majk09a,Majk09f,Majk08mm,Majk11mm,MaBh10} as well, where this basic power-view operator $T$ is extended to the
endofunctor $T:\textbf{DB} \rightarrow \textbf{DB}$.\\
The connection between a logical (schema) level and this
computational category is based on the \emph{interpretation}
functors. Thus, each rule-based conjunctive query at schema level
over a database $\A$ will be translated (by an interpretation
functor) in a morphism in $\textbf{DB}$,  from an instance-database
$A$ (a model of the database schema $\A$) to the instance-database
$TA$ composed by all views of  $A$.

\subsection{Basic Database concepts}
The database mappings,  for a given logical language (for default we
assume the First-Order Language (FOL)), are defined usually at a
schema level ($\pi_1,\pi_2$ denote first and second projections,
$\biguplus$ disjoint union, and $\mathcal{N}$ the set of natural
numbers), as follows:
\begin{itemize}
  \item  A \emph{database schema} is a pair $\A = (S_A , \Sigma_A)$ where: $S_A = \pi_1(\A)$ is
  a countable set of relation symbols $r\in R, ~ar:R \rightarrow \mathcal{N}$, with finite arity
  (finite list of attributes $\textbf{x} = <x_1,...,x_n>, n = ar(r) \geq 1$),
disjoint from a countable infinite set $\textbf{att}$ of attributes
(for any single attribute $x \in \textbf{att}$ a domain of $x$ is a
 nonempty subset $dom(x)$ of a countable set of individual symbols
$\textbf{dom}$, disjoint from $\textbf{att}$ ), such that for any
$r\in R$, the sort of $R$ is a finite sequence of elements of
$\textbf{att}$. $\Sigma_A = \pi_2(\A)$ denotes a set of closed
formulas (without free variables) called integrity constraints, of
the sorted First-Order Language (FOL) with sorts $\textbf{att}$,
constant symbols $\textbf{dom}$,
relational symbols in $S_A$, and no function symbols.\\
We denote by  $\mathbb{S}$ the set of all
database schemas for a given (also infinite) set $R$.\\
We denote by $\A_\emptyset$ the empty database schema (where
$\pi_1(\A_\emptyset)$  and $\pi_2(\A_\emptyset)$ are empty sets). A
\emph{finite} database schema $\A$ is composed by a finite set
$S_A$, so that the set of all attributes of such a database is
finite.
\item We consider a rule-based
\emph{conjunctive query} over a database schema $\A$ as an
expression $ q(\textbf{x})\longleftarrow R_1(\textbf{u}_1), ...,
R_n(\textbf{u}_n)$, where $n\geq 0$, $R_i$ are the relation names
(at least one) in $\A$ or the built-in predicates (ex. $\leq, =,$
etc..), $q$ is a relation name not in $\A$, $\textbf{u}_i$ are free
tuples (i.e., may use either variables or constants). Recall that if
$\textbf{v} = (v_1,..,v_m)$ then $R(\textbf{v})$ is a shorthand for
$R(v_1,..,v_m)$. Finally, each variable occurring in $\textbf{x}$
must also occur at least once in $\textbf{u}_1,...,\textbf{u}_n$.
Rule-based conjunctive queries (called rules) are composed by: a
subexpression $R_1(\textbf{u}_1) , ...., R_n(\textbf{u}_n)$, that is
the \emph{body}, and $q(\textbf{x})$ that is the \emph{head} of this
rule. If one can find values for the variables of the rule, such
that the body holds (i.e. is logically satisfied), then one may
deduce the head-fact.
 This concept is captured by a notion of "valuation". In the
 rest of this paper a deduced head-fact will be called  "a resulting \emph{view}
 of a query $q(\textbf{x})$ defined over a database $\A$", and denoted by $\|q(\textbf{x})\|$. Recall that the conjunctive
queries are monotonic and satisfiable. The $Yes/No$ conjunctive
queries are the rules with an empty head.
 \item We consider that a
\emph{mapping} between two database schemas $\A$ and $\B$ is
expressed by an union of "conjunctive queries with the same head".
Such mappings are called "view-based mappings" and can be defined by
a set $~\M = \{q_{Ai}(\textbf{x}_i) \Rightarrow q_{Bi}(\textbf{x}_i)
| 1 \leq i \leq n \}$, where $\Rightarrow$ is the logic implication
between these conjunctive queries
$q_{Ai}(\textbf{x}_i)$ and $q_{Ai}(\textbf{x}_i)$, over databases $\A$ and $\B$ respectively.\\
% where  \emph{a view} is defined bysuch an union of conjunctive queries.
 We consider \emph{a
view} of an instance-database $A$ an n-ary relation (set of tuples)
obtained by a "select-project-join + union" (SPJRU) query
$q(\textbf{x})$  (it is a term of SPJRU algebra)  over
    $A$: if this query is a finite term of this algebra than it is
    called a "finitary view" (a finitary view can have also an infinite number of
    tuples).
\item An \emph{instance} of a database $\A$ is given by $A = (\A,I_A)$, where
$I_A$ is an Tarski's FOL interpretation function, that satisfies
\emph{all} integrity constraints in  $\Sigma_A$, and maps each
relational symbol of $S_A$ (n-ary predicate in FOL) into an n-ary
relation $a_i\in A $ (called also "element of  $A$" ). Thus, a
relational instance-database $A$ is a set of n-ary relations, and
they are managed by  relational database systems (RDBMS).\\
 Given
two autonomous instance-databases $A$ and $B$, we can make a
\emph{federation} of them, i.e., their disjoint union $A \biguplus
B$, in order to be able to \emph{compute} the queries with relations
of
both autonomous instance-databases.\\
A federated database system is a type of meta-database management
system (DBMS)  which transparently integrates multiple autonomous
database systems into a single federated database.
  The constituent databases are interconnected via a computer network, and may be geographically decentralized.
  Since the constituent database systems remain autonomous, a federated database system is a contrastable alternative to the
  (sometimes daunting) task of \emph{merging} together several disparate databases. A federated database, or virtual database,
  is the fully-integrated, logical composite of all constituent databases in a federated database
  system.\\
  McLeod and Heimbigner \cite{ShLa90} were among the first to define a federated database system,
  as one which "define[s] the architecture and interconnect[s] databases that minimize central authority yet support partial
  sharing and coordination among database systems".
Among other surveys, Sheth and Larsen \cite{LeHe85} define a
Federated Database as a collection of cooperating component systems
which are autonomous and are possibly heterogeneous.
 \end{itemize}
\subsection{DB (Database) category \label{Section:DB}} Based on an
observational point of view for relational databases, we may
introduce a category $\textbf{DB}$ \cite{Majk04AOT} for
instance-databases and view-based mappings between them, with the
set of its objects $Ob_{DB}$, and the set of its morphisms
$Mor_{DB}$, such that:
\begin{enumerate}
  \item Every object  (denoted by $A,B,C$,..) of this category is a
  instance-database, composed by a set of n-ary relations  $a_i\in A$, $i= 1,2,...$  called also "elements of
  $A$".
 We define a universal database instance  $\Upsilon$
  as the  union of all database instances, i.e., $\Upsilon = \{ a_i | a_i\in A, A\in Ob_{DB}\}$.
  It  is the top object of this category.\\
     We have that $\Upsilon = T\Upsilon$, because every view $v\in T\Upsilon$
  is  an instance-database as well, thus $v\in \Upsilon$. Vice versa, every element
   $r\in \Upsilon$ is  a view of $\Upsilon$ as well, thus $r\in T\Upsilon$.\\
      Every object (instance-database) $A$ has also the empty relation $\bot$. The object  composed by only this
  empty relation is denoted by $\bot^0$ and we have that $T\bot^0
  =\bot^0= \{\bot\}$.\\
  Two objects $A$ and $B$ are isomorphic in $\textbf{DB}$, denoted
  by $A \simeq B$, if $TA = TB$.\\
  For any instance-database $A$ it holds that $A \subseteq TA$ and $A \simeq TA$.\\
  Any empty database (a database with only empty relations) is isomorphic to this bottom object
  $\bot^0$.
  \item Morphisms of this category are all possible mappings
  between instance-databases \emph{based on views}, as they will be defined
  by formalism of operads in what follows.\\
  \end{enumerate}
In what follows, the objects in $\textbf{DB}$ (i.e.,
instance-databases) will be called simply databases as well, when it
is clear from the context.
%\subsection{Typed operads and schema mappings}
% Now we are ready to formally define the set "c-arrow" of all \emph{complete
%morphisms} for databases.
Each atomic mapping (morphism) in $\textbf{DB}$ between two
databases is generally composed of three components: the first
correspond to conjunctive query $q_i$ over a source database that
defines this view-based mapping, the second (optional) $w_i$
"translate" the obtained tuples from domain of the source database
(for example in Italian) into terms of domain of the target database
(for example in English), and the last component $v_i$ defines which
contribution of this mappings is given to the target relation, i.e.,
a kind of Global-or-Local-As-View (GLAV) mapping
(sound, complete or exact).\\
In what follows we will consider more simple case without the
component $w_i$.\\
%\begin{definition}\label{def:view-map} \textsc{View-mapping:}
%  For any \textsl{query} over a  schema $\A$  we can define a
%  schema map $q_i:\A\longrightarrow T\A $,  where $q_i\in O(r_{i1},..., r_{ik}, r_i )$,
%  $~Q = (r_{i1},..., r_{ik})\subseteq~\A$,  and $r_i\in T\A$.\\  A correspondent view-map
%  at instance level is $q_{A_i}= \{\alpha(q_i), q_\perp\}:A \longrightarrow TA$, with $A = \alpha^*(\A)$,
%  $TA = \alpha^*(T\A)$, $~\partial_0(q_\perp)= \partial_1(q_\perp)=
%  \{\perp\}$. For simplicity, in the rest we will drop the
%  component $q_\perp~$ of a view-map, and implicitly assume such
%  component; thus
%    $~\partial_0(q_{A_i}) = \alpha^*(Q)\subseteq A ~$ and $~\partial_1(q_{A_i})= \{\alpha(r)\} \subseteq TA
%  ~$ is the view obtained by this "select-project-join+union"
%  query $~q_i$.
%\end{definition}
%Instead of lists $(g_1,...,g_k)$ used
%  for mappings in Definitions \ref{def:operads}, \ref{def:operads2}, we will use the sets
%  $\{g_1,...,g_k\}$ because a mapping between two databases does not
%  depend on a particular permutation of its components.
We introduce also the two functions $\partial_0, \partial_1$ such
that $\partial_0(q_{A_i}) = \{r_{i1},...,r_{ik}\}$ (the set of
relations used in the query formula $q_{A_i}(\textbf{x})$ and
$\partial_1(q_{A_i})
= \{r_i\}$, with obtained  view $ r_i = \|q_{A_i}(\textbf{x})\|$.\\
Thus, we can formally introduce a theory for view-mapings based on
operads:
%at instance level, which is an component of the attribute mapping $~~Out$, as will be formally specified by operads as follows:
\begin{definition}\label{def:view-map}
%We denote by $\A_\varnothing$ the  database shema with only
%a special relation symbol $r_\varnothing$ such that for each
% $\alpha$ it holds that $\alpha(r_\varnothing) = \perp $, where
%$\perp$ denotes the empty relation.
We define the following two types of basic mappings:
  \begin{itemize}
  \item \textsc{logic-sentence mapping}:
  %for any Yes/No query, or
  For any  sentence, a logic formula $\varphi_i$  without free variables  over a
  Database schema $\A$,  we can define a
  schema mapping $~\varphi_i:\A\longrightarrow \A_\varnothing $.
  The unique instance-database  of the empty shema $\A_\varnothing$ is
denoted by $\perp^0 = \{\perp \}$, where $\perp$ denotes the empty
relation. Consequently, for each interpretation $\alpha$ it holds
that $\alpha^*(\A_\emptyset) = \perp^0$.\\
  This kind of schema mappings will be used for the integrity
  constraints over database schemas, and Yes/No queries, as will be
  specified in Section \ref{Section:interpret}.
%  ,  where $q_i\in O(r_{i1},..., r_{ik}, r_\varnothing )$,
%  $~Q = (r_{i1},..., r_{ik})\subseteq~\A$.\\  For a given  $\alpha$,  such that $q_i$ is true, the correspondent view-map at instance level is
%   $q_{A_i}=  \alpha(q_i):A \longrightarrow \perp^0$.
   %
    \item   \textsc{View-mapping}:
  For any  \textsl{query} (a logic formula with free variables) over a  schema $\A$  we can define a
  schema map $q_i:\A\longrightarrow \{r_i\} $,  where $q_i\in O(r_{i1},..., r_{ik}, r_i )$,
  $~Q = (r_{i1},..., r_{ik})\subseteq~\A$.\\  For a given $\alpha$ the correspondent view-map
  at instance level is $q_{A_i}= \{\alpha(q_i), q_\perp\}:A \longrightarrow TA$, with $\perp \in A = \alpha^*(\A)\subseteq TA)$, $~\partial_0(q_\perp)= \partial_1(q_\perp)=
  \{\perp\}$. For simplicity, in the rest of this paper we will drop the
  component $q_\perp~$ of a view-map, and assume implicitly  such a
  component; thus,
    $~\partial_0(q_{A_i}) = \alpha^*(Q)\subseteq A ~$ and $~\partial_1(q_{A_i})= \{\alpha(r_i)\} \subseteq TA
  ~$ is a singleton with the unique element equal to view obtained by a "select-project-join+union"
  term $~\widehat{q_i}$.
    \end{itemize}
\end{definition}
Thus, we introduce an \emph{atomic morphism} (mapping) between two
databases as a set of simple view-mappings:
%In what follows, for
%generality, such view-mappings will be enriched by two other
%components: one for "translations" of terms of source database into
%terms of target database (for example if they are given in two
%different languages), and second one for a kind of GLAV mapping
%(sound, complete or exact).
\begin{definition}\label{def:atomicmorphisms} \textsc{Atomic morphism:}
%Lets define:
%\begin{enumerate}
%  \item for a \textsl{query-mapping} over a  schema $\A$  we can define a
%  schema map $q_i:\A\longrightarrow T\A $,  where $q_i\in O(r_{i1},..., r_{ik}, r_i )$,
%  $~Q = (r_{i1},..., r_{ik})\subseteq~\A$,  and $r_i\in T\A$.  A correspondent view-map
%  at instance level is $q_{A_i}= \{\alpha(q_i), q_\perp\}:A \longrightarrow TA$, with $A = \alpha^*(\A)$,
%  $TA = \alpha^*(T\A)$, $~\partial_0(q_\perp)= \partial_1(q_\perp)=
%  \{\perp\}$. For simplicity, in the rest we will drop the
%  component $q_\perp~$ of a view-map, and implicitly assume such
%  component; thus
%    $~\partial_0(q_{A_i}) = \alpha^*(Q)\subseteq A ~$ and $~\partial_1(q_{A_i})= \{\alpha(r)\} \subseteq TA
%  ~$ is the view obtained by this "select-project-join+union"
%  query $~q_i$.
%  \item
Every \textsl{schema mapping} $~f_{Sch}:\A\longrightarrow \B$, based
on a set of query-mappings $q_i$, is defined
  for finite natural number N by\\
  $ f_{Sch} \triangleq \{~v_i\cdot  q_i~ |~ q_i\in O(r_{i1},..., r_{ik},~
  r_i'),  ~v_i\in O(r_i',r_i),\\~ \{r_{i1},..., r_{ik}\}\subseteq \A ,~ r_i\in
  \B, 1\leq i\leq N \}$.\\
Its correspondent \textsl{complete morphism} at instance database level is\\
  $f = \alpha^*(f_{Sch}) \triangleq \{~q_{A_i} = \alpha(v_i)\cdot  \alpha(q_i)~|~
  ~ v_i\cdot  q_i \in f_{Sch}\}:A \rightarrow B$,
  % also   denoted by (for simplicity we write $q_{A_i}$ instead  $\{q_{A_i}\}$)
 %$ ~~~~f =  \bigcup_{ q_{A_i}\in S_f} q_{A_i}:A\longrightarrow B$, with $S_f = \alpha^*(f_{Sch})$,
where:\\
  Each $\alpha (q_i)$ is a query computation, with obtained view $\alpha(r_i') \in TA$
  for an instance-database  $A = \alpha^*(\A) = \{ \alpha(r_k)~|~r_k \in \A \}$, and $B = \alpha^*(\B)$.\\
%  Each $\alpha(w_i):\alpha(r_i'')\longrightarrow \alpha(r_i')$, where $\alpha(r_i') \in TB$, is
% equal to the function determined by the symmetric \textsl{domain
% relation} $~r_{AB}\subseteq \textbf{dom}_A \times \textbf{dom}_B~$
% for the equivalent constants in $\alpha^*(\A)~and~\alpha^*(\B)~$ ($(a,b)\in r_{AB}~$ means that, $a \in \textbf{dom}_A~and~b \in
% \textbf{dom}_B~$ represent the same entity of the real word (requested for a federated database environment) as:
% for any $(a_1,...,a_n) \in \alpha(r'')~holds~
% \alpha(w_i)(a_1,...,a_n)=(b_1,...,b_n)$, and for all $1\leq
% k\leq n ~~(a_k,b_k) \in r_{AB}$.
% If $r_{AB}~$ is not defined, it is assumed, by default,  that
% $\alpha(w_i)~$ is an identity function.\\
 Let $\pi_{q_i}$ be a projection function on relations, for all attributes in
 $\partial_1(\alpha(q_i)) = \{\alpha(r_i'')\}$. Then,
  each $\alpha (v_i):\alpha(r_i')\longrightarrow \alpha (r_i)$ is one  tuple-mapping function,
   used to distinguish sound
%   ,   complete
   and exact assumptions on the views, as follows:
\begin{enumerate}
  \item \textsl{inclusion} case, when $~ \alpha(r_i')\subseteq \pi_{q_i}(\alpha
  (r_i))$. Then
    for any tuple $t\in \alpha(r_i')$, $~~\alpha (v_i)(t)
  = t_1$, for some $t_1\in\alpha(r_i)~$ such that $\pi_{q_i}(\{t_1\})=
  t$.
%  \\ We define  $\|q_{A_i} \| \triangleq  \alpha(r_i')$ the extension of data
%  transmitted from an instance-database $A$ into $B$ by the component $q_{A_i}$.
%
%  \item \textsl{inverse-inclusion} case, when $ \alpha(r_i')\supseteq \pi_{q_i}(\alpha
%  (r_i))$.\\
%  Then, for any tuple $t\in \alpha(r_i')$,
%  \begin{displaymath}
%   \alpha (v_i)(t) = \left\{ \begin{array}{ll}
%   t_1 & \textrm{ ~, ~if ~$ \exists{t_1}\in \alpha (r_i)~,~\pi_{q_i}(\{t_1\})
%  = t$}\\
%& \textrm{~empty ~tuple, ~otherwise}
%  \end{array} \right.
%  \end{displaymath}
%  We define  $\|q_{A_i} \| \triangleq  \pi_{q_i}(\alpha
%  (r_i))$ the extension of data
%  transmitted from an instance-database  $A$ into $B$ by the component $q_{A_i}$.
%
%  \item \textsl{equal} case, when both (a) and (b) are valid.
  \item \textsl{exact} case, special inclusion case when $~ \alpha(r_i')= \pi_{q_i}(\alpha
  (r_i))$.
  \end{enumerate}
   We define  $\|q_{A_i} \| \triangleq  \alpha(r_i')$ the extension of data
  transmitted from an instance-database $A$ into $B$ by the component $q_{A_i}$.
%   \end{enumerate}
     \end{definition}
  Notice that the components $\alpha(v_i),  \alpha(q_i)$ are
  not the morphisms in $\textbf{DB}$ category:  only their functional
  composition is an atomic morphism.
  Each atomic morphism is a complete morphism, that is, a set of view-mappings. Thus,
  each view-map $~q_{A_i}:A\longrightarrow TA$, which is an atomic morphism,
   is a complete morphism (the case when $B = TA$,  and
   $\alpha (v_i)$ belongs to the "exact case"),
  and  by \emph{c-arrow} we denote the set of all complete
  morphisms.\\
  Based on atomic morphisms (sets of view-mappings) which are complete
arrows (c-arrows), we obtain that their composition generates
tree-structures, which can be incomplete (p-arrows), in the way that
for a composed arrow $h = g\circ f:A \rightarrow C$, of two atomic
arrows $f:A \rightarrow B$ and $g:B\rightarrow C$, we can have the
situations where $\partial_0(f) \subset
\partial_0(h)$, where the set of relations in $\partial_0(h) -
\partial_0(f) \subset
\partial_0(g)$ are denominated "hidden elements".
\begin{definition}\label{def:morphisms}  The following BNF defines the set $Mor_{DB}$ of all morphisms in
DB:\\
  $~~~p-arrow \textbf~{:=} ~c-arrow ~|~c-arrow \circ c-arrow~$ (for any two c-arrows  $f:A\longrightarrow B$ and $g:B\longrightarrow C~$)\\
 $~~~morphism \textbf~{:=} ~p-arrow ~|~c-arrow \circ p-arrow~$ (for any p-arrow $f:A\longrightarrow B$ and c-arrow $g:B\longrightarrow C$)\\
\\whereby the composition of two arrows, f (partial) and g (complete),
we obtain the following p-arrow (partial arrow) $h = g \circ
f:A\longrightarrow C$
\[
\ h = g\circ f = \bigcup_{ q_{B_j}\in ~g ~\& ~\partial_0 (q_{B_j})
\bigcap
\partial_1 (f) \neq \emptyset } \{q_{B_j}\}~~~~\circ
\]
\[
\circ ~~~~ \bigcup_{q_{A_i}\in ~f ~\&~\partial_1 (q_{A_i})=
\{v\}~\&~ v \in ~\partial_0 (q_{B_j}) } \{q_{A_i}(tree)\}~~
\]
$= \{q_{B_j} \circ \{q_{A_i}(tree)~|~\partial_1(q_{A_i}) \subseteq
\partial_0(q_{B_j})\}~|~q_{B_j}\in ~g ~\&
~\partial_0 (q_{B_j}) \bigcap
\partial_1 (f) \neq \emptyset \}$\\
$= \{q_{B_j}(tree)~|~q_{B_j}\in ~g ~\& ~\partial_0 (q_{B_j}) \bigcap
\partial_1 (f) \neq \emptyset\}$
\\where   $q_{A_i}(tree)$ is the  tree of the morphisms f below
$q_{A_i}$.\\
We define the semantics of mappings by function
 $B_T:Mor_{DB}\longrightarrow Ob_{DB}$, which, given any mapping
 morphism
 $f:A\longrightarrow B~$ returns with the set of views ("information flux") which are
 really "transmitted" from the source to the target object.\\
 1. for atomic morphism, $ \widetilde{f} = B_T(f)~\triangleq
 T\{\|f_i\|~|~f_i \in f \}$.\\
 2. Let $g:A \rightarrow B$ be a morphism with a flux
 $\widetilde{g}$, and $f:B \rightarrow C$ an atomic morphism with
 flux $\widetilde{f}$ defined in point 1, then  $~~\widetilde{f \circ g}
 = B_T(f \circ g) ~\triangleq \widetilde{f}\bigcap
 %T\{v ~|~ v \in \partial_0(f) \bigcap  \partial_1(g)\} \bigcap
 \widetilde{g}$.\\
 We introduce an equivalence relation over
morphisms by, $ ~~~~f \approx g ~~~~iff~~~~ \widetilde{f} =
\widetilde{g}$.
 \end{definition}
%\]
Notice that between any two databases $A$ and $B$ there is at least
an "empty" arrow $\emptyset:A \rightarrow B$ such that
$\partial_0(\emptyset) =
\partial_1(\emptyset) = \widetilde{\emptyset} = \{\bot\} =
\bot^0$.\\
  Basic properties of this
database category $\textbf{DB}$ as its symmetry (bijective
correspondence between arrows and objects, duality ($\textbf{DB}$ is
equal to its dual $\textbf{DB}^{OP}$) so that each limit is also
colimit (ex. product is also coproduct, pullback is also pushout,
empty database $\bot^0$ is zero objet, that is, both initial and
terminal object, etc..), and that it is a 2-category has been
demonstrated in
\cite{Majk03,Majk04AOT}.\\
Generally, database mappings are not simply programs from values
(relations) into computations (views) but an equivalence of
computations: because of that each mapping, from any two databases
$A$ and $B$, is symmetric and gives a duality property to the
category $\textbf{DB}$. The denotational semantics of database
mappings is given by morphisms of the Kleisli category
$\textbf{DB}_T$ which may be "internalized"
in $\textbf{DB}$ category as "computations" \cite{MaBh10}.\\
 The product $A\times B$ of a databases
$A$ and $B$ is equal to their coproduct $A+B$, and the semantics for
them is that we are not able to define a view by using relations of
both databases, that is, these two databases have independent DBMS
for query evaluation. For example, the creation of exact copy of a
database $A$ in another DB server corresponds to the database $A +
A$. \\
The duality property for products and coproducts are given by the
following commutative diagram:
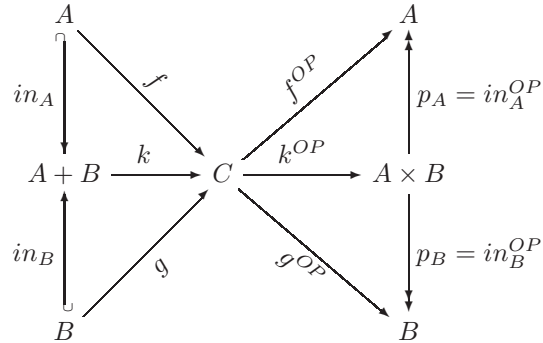
\begin{diagram}
A            &          &   &                 &     A           \\
\dInto^{in_A}  & \rdTo^f  &   & \ruTo^{f^{OP}} &\uOnto_{p_A = in_A^{OP}}\\
A+B          & \rTo^k   & C & \rTo^{k^{OP}}  & A\times B  \\
\uInto^{in_B}  &  \ruTo_g &   & \rdTo_{g^{OP}} &  \dOnto_{p_B = in_B^{OP}}\\
 B           &          &   &                 &   B            \\
\end{diagram}
In the paper \cite{Majk09a,Majk09f,Majk09a} have been considered
some relationships of $\textbf{DB}$ and standard $\textbf{Set}$
category, and has been introduced the categorial (functors)
semantics for two basic database operations: \emph{matching}
$\otimes$,  and \emph{merging} $\oplus$, such that for any two
databases $A$ and $B$, we have that $A \otimes B = TA \bigcap TB$
and $A \oplus B = T(A \bigcup B)$. In the same work has been defined
the algebraic database lattice and has been shown that $\textbf{DB}$
is concrete, small and locally finitely presentable (lfp) category.
Moreover, it was shown that $DB$ is also V-category enriched over
itself, was developed a metric space and a subobject classifier for
this
category, and demonstrated that it is a weak monoidal topos.\\
In this paper we will develop a functorial semantics for the database schema mapping system, based on the theory of sketches. \\
%\
%

 The plan of this paper is the following: In Section 2 we will present
 the two basic operations for database schemas used in database
 mapping systems (separation and federation), and we will explain
 why the functorial semantics for database mappings needed a new
 base category instead of common $\textbf{Set}$ category.\\
 Finally, in Section 3 is presented  a definition of the graph $G$ for a schema database
 mapping system, and the definition of its sketch category
 $\textbf{Sch}(G)$. Based on this framework then is presented
 functorial semantics for database mapping systems with the base
 category $\textbf{DB}$.
\section{Basic database schema operations: Separation and
Federation \label{section:prelim}}
For the composition of complex database mapping graphs, it is
important to distinguish two basic compositions of two database
schemas $\A$ and $\B$ with respect to DBMSs:
\begin{itemize}
  \item the case when in this composed schema,  the two  database schemas are mutually
  \emph{separated} by two independent DBMSs, in order that it is
  impossible to write a query over this composition with relations of both databases: it
  is common case when two databases are separated, and this
  symmetric binary  separation-composition at schema level will be
  denoted by $\A \dagger \B$, such that $\pi_i(\A \dagger \B) = \pi_i(\A) \biguplus \pi_i(\B), i = 1,2$.
  \item the case when in this composed schema, the two  database schemas are \emph{connected} into the same DBMS (without any change of the two original
  database schemas):
  In this case we are able to use the queries over this composed schema with relations \emph{of both}
  databases for inter database mappings, and this
  symmetric binary  federation-composition at schema level will be
  denoted by  $\A  \bigoplus \B$,  such that $\pi_i(\A \bigoplus \B) = \pi_i(\A) \biguplus \pi_i(\B), i = 1,2$.
\end{itemize}
The identity $=$ for  database schemas is naturally defined by: for
any two $\A, \B \in \mathbb{S}$, $\A = \B$ if $\pi_i(\A) =
\pi_i(\B), i = 1,2$.  Notice that both symmetric binary operators,
$\dagger, \bigoplus$, for  database schemas in $\mathbb{S}$ are
associative with identity element $\A_\emptyset)$ (nullary
operator), so that the algebraic structures $((\mathbb{S}, =),
\dagger,\A_\emptyset)$
and $((\mathbb{S}, =), \bigoplus,\A_\emptyset)$ are the monoids.\\
%, with the distribution low: $\A \bigoplus(\B \dagger \C) = (\A \bigoplus \B) \dagger (\A \bigoplus \C)$.\\
 Let us consider, for example, a
mapping $\M:\A \dagger \B \rightarrow \C$, and a mapping $\M:\A
\bigoplus \B \rightarrow \C$. In the first case in any query mapping
$q(\textbf{x}) \Rightarrow q_C(\textbf{x}) \in \M$ the all relation
symbols in the query $q(\textbf{x})$ must be of database $\A$ or
(mutually exclusive) of database $\B$, and this mapping can be
equivalently represent by the graph:
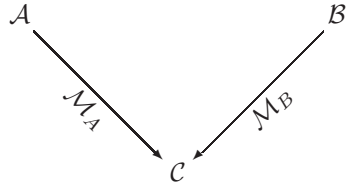
\begin{diagram}
 \A        &           &       &    &\B \\
           & \rdTo_{\M_A}&      &  \ldTo_{\M_B} &  \\
           &           &   \C   &    &\\
\end{diagram}
where $\M_A \biguplus \M_B = \M$,  while in the case of mapping
$\M:\A \bigoplus \B \rightarrow \C$ such an decomposition is not
possible, because we can have a query mapping $q(\textbf{x})
\Rightarrow q_C(\textbf{x}) \in \M$ with relation symbols from both
databases
$\A$ and $\B$.\\
If we introduce the mappings $\M_1 = \{ r_{Ai}(\textbf{x}_i)
\Rightarrow r_{Ai}(\textbf{x}_i)| r_{Ai} \in \A \}$ and  $\M_2 =
\{r_{Bi}(\textbf{y}_i) \Rightarrow r_{Bi}(\textbf{y}_i)| r_{Bi} \in
\B \}$, then we obtain the  mapping graph,
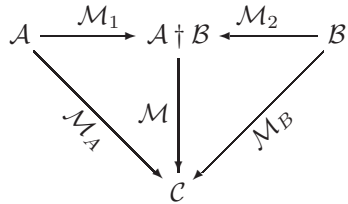
\begin{diagram}
 \A        & \rTo{\M_1}          &  \A \dagger \B     &  \lTo{\M_2}    & \B \\
           & \rdTo_{\M_A}        &   \dTo{\M}               &  \ldTo_{\M_B}  &  \\
           &                     &   \C             &                & \\
\end{diagram}
%\begin{center}
% \textbf{Fig.A} . Cocone diagram
% \end{center}
that can be seen as a cocone diagram for schema database mappings.\\
Let us consider another dual example, a mapping $\M:\C \rightarrow
\A \dagger \B$. In this case in any query mapping $q_C(\textbf{x})
\Rightarrow q(\textbf{x}) \in \M$ the all relation symbols in the
query $q(\textbf{x})$ must be of database $\A$ or (mutually
exclusive) of database $\B$, and this mapping can be equivalently
represent by the graph:
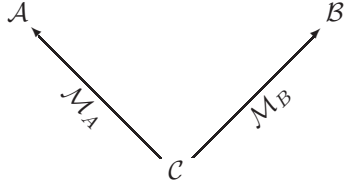
\begin{diagram}
 \A        &           &       &    &\B \\
           & \luTo_{\M_A}&      &  \ruTo_{\M_B} &  \\
           &           &   \C   &    &\\
\end{diagram}
where $\M_A \biguplus \M_B = \M$.\\
If we again introduce the mappings $\M_1 = \{ r_{Ai}(\textbf{x}_i)
\Rightarrow r_{Ai}(\textbf{x}_i)| r_{Ai} \in \A \}$ and  $\M_2 =
\{r_{Bi}(\textbf{y}_i) \Rightarrow r_{Bi}(\textbf{y}_i)| r_{Bi} \in
\B \}$, then we obtain  dual mapping graph,
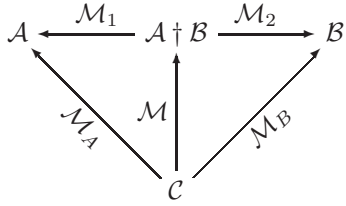
\begin{diagram}
 \A        & \lTo{\M_1}          &  \A \dagger \B     &  \rTo{\M_2}    & \B \\
           & \luTo_{\M_A}        &   \uTo{\M}               &  \ruTo_{\M_B}  &  \\
           &                     &   \C             &                & \\
\end{diagram}
%\begin{center}
% \textbf{Fig.B}. Cone diagram
% \end{center}
that can be seen as a cone diagram for schema database mappings.\\

Based on these two simple examples, generally, the schema database
mappings can be expressed by using the small \emph{sketches}. The
detailed presentation of sketches
 for the database mappings, and their functorial semantics will be given in Section \ref{Section:interpret}.\\
Sketches are developed by Ehresmann's school, especially by
R.Guitartand and C.Lair \cite{Ehre66,Liar96,BaWe88}. Sketch is a
category together with distinguished class of cones and cocones. A
\emph{model} of the sketch is a set-valued functor turning all
distinguished cones into limit cones, and all distinguished cocones
into colimit cocones, in
the category  $\textbf{Set}$ of sets.\\
There is an elementary and basic connection between sketches and
logic. Given any sketch, one can consider the underlying graph of
the sketch as a (many-sorted) language, and one can write down
axioms in the $\L_{\infty,\infty}$-logic (the infinitary FOL with
finite quantifiers) over this language, so that the models of the
axioms become exactly the models of the sketch.\\
the category of models of a given sketch has as objects the models,
and arrows all natural transformations between the models as
functors. A category is sketchable (esquissable) or accessible iff
it is equivalent to the category of set-valued models of a small
sketch.\\
Recall that a graph $G$ consists of a set of vertices denoted $G_0$
and a set of arrows denoted $G_1$ together with the operators $dom,
cod:G_1 \rightarrow G_0$ which assigns to each arrow its source and
target. (Co)cones and diagrams are defined for graphs in exactly the
same way as they are for categories, but commutative co(cones) and
diagrams of course make no sense for graphs.\\
By a sketch me mean 4-tuple $(G,u,D,C)$ where $G$ is a graph,
$u:G_0\rightarrow G_1$ is a function which takes each vertex (node)
$\A$ in $G_0$ to an arrow from $\A$ to $\A$, $D$ is a class of
diagrams in $G$ and $C$ is class of (co)cones in $G$. Each (c)cone
in $G$ goes (to)from some vertex (from)to some diagram; that diagram
need not be in $D$, in fact it is necessary to allow diagrams which
are not in $D$ as bases of (co)cones.\\
Notice that, differently from the work dedicated to categorical
semantics of Entity-Relationship internal relational database
models, where nodes of sketches are single relations, here at higher
level of abstraction, the nodes are whole databases: consequently in
a such framework we do not use commutative database mapping systems,
thus $D$ is empty set. In fact, in the database mapping system, the
(co)cone diagrams above will never be used in practical
representation of database mapping systems: instead will be
alternatively used only its selfconsistent parts, as first diagram
above, or, equivalently, a single arrow $\M:\A \dagger \B
\rightarrow
\C$.\\
But, clearly, for the introduced dataschema composition operator
$\dagger$, with cone and cocone diagrams above has to be
present in $C$ for our sketches.\\
Consequently we obtain the following fundamental lemma for the
categorial modelling of database mappings:
\begin{lemma} \label{lemma:basecategory} The \textbf{Set}  can not be used as the base category for the models of
database-mapping sketches.
\end{lemma}
\textbf{Proof}: Let $\textbf{E}$ be a sketch for a given database
sketch $(G,u,D,C)$, where $C$ is a set of (co)cones of the two
diagrams introduced for the database schema composition operator
$\dagger$, and a model of this sketch be a functor $F:\textbf{E}
\rightarrow \mathbb{B}$, where $\mathbb{B}$ is a base category. Then
all cones in $C$ has to be functorially translated into limit
commutative diagrams in $\mathbb{B}$, and all cocones in $C$ has to
be functorially translated into limit commutative diagrams in
$\mathbb{B}$: i.e., the cocone in the figure above has to be
translated into coproduct diagram, and cone into product diagram in
$\mathbb{B}$.\\
Consequently, the object $F(\A \dagger \B)$ has to be both the
product $A \times B$ and coproduct $A + B$, where $A = F(\A)$ and $B
= F(\B)$ are two objects in $\mathbb{B}$ and $\times, +$ the product
and coproduct operators in $\mathbb{B}$, but it can not be done in
\textbf{Set}. In fact, the product $A \times B$ in \textbf{Set} is
the cartesian product of these two sets $A$ and $B$, while the
coproduct $A + B$ is the disjoint union, so that does not hold the
isomorphism $A \times B ~\simeq~ A + B$ in \textbf{Set}.
\\$\square$\\
\textbf{Remark}: The fundamental consequence of this lemma is that
we needed to define a new base category for the categorial semantics
of database mappings. In fact, we  defined this new base category
$\mathbb{B}$, denoted by $\textbf{DB}$ (DataBase) category, and we
have  shown that it satisfies  the duality property where the
product and coproduct diagrams are dual diagrams, so that for any
two objects (instance databases) in $\textbf{DB}$ the objects $A
\times B$ and $A + B$ are equal (up to isomorphism).

%%%%%%%%%%%%%%%%%%%%%%%%%%%%%%%%%%%%%%%%%%%%%%%%%%%%%%%%
%
%
\subsection{Data Separation: (Co)product  operator in DB}
Separation-composition of objects are coproducts (and products) in
$\textbf{DB}$ category:
\begin{definition} \label{def:coproduct}
The disjoint union  of any two instance-databases (objects) $A$ and
$B$, denoted by $~A+B$, corresponds to two mutually isolated
databases, where two database management systems are completely
disjoint, so that it is impossible to compute the queries with
the relations from both databases.\\
The disjoint property for mappings is represented by facts that\\
$\partial_0(f+g)\triangleq \partial_0(f)+
\partial_0(g),~~~\partial_1(f+g)\triangleq \partial_1(f)+
\partial_1(g)$.
\end{definition}
Thus, for any database $A$, the \emph{replication} of this database
(over different DB servers) can be denoted by the coproduct object
$A + A$ in this category $\textbf{DB}$.
\begin{propo}\label{prop:disjiont}
For any two databases (objects) $A$ and $B$ we have that $T(A+B)= TA
+  TB$. Consequently $A + A$ is not isomorphic to $A$.
% The following properties for the disjoint union
% of the definition above are valid:
%\begin{enumerate}
%  \item For any two databases (objects) A and B we have that $T(A+B)= TA +
%  TB$, and
%    for any two arrows f,g we have that $~~\widetilde{f+g} = \widetilde{f} +
%  \widetilde{g}$.
%    \item For any object C in DB
%   $~~ C + \perp^0 ~=~  \perp^0 + C ~\simeq ~C$.
%  \item For any arrow f in DB
%  $ ~~f + \perp^1 ~\approx~ \perp^1 + f ~\approx~
%  f$, where $~\perp^1~$ is banal empty morphism between objects, such that
%  $~\partial_0(\perp^1)= \partial_1(\perp^1) = \perp^0$, with $\widetilde{\perp^1} = \perp^0$.
%\end{enumerate}
\end{propo}
%\begin{proof}
\textbf{Proof:}  We have that $ T(A+B)= TA +
  TB$, directly from the fact that we are  able to define views
  only over relations in $A$ or, alternatively, over relations in $B$. Analogously $ ~~\widetilde{f+g} = \widetilde{f} +
  \widetilde{g}$, which is a closed object, that is, holds that $T(\widetilde{f+g}) = T(\widetilde{f} +
  \widetilde{g}) = T\widetilde{f} +
  T\widetilde{g} = \widetilde{f} +
  \widetilde{g} = \widetilde{f+g}$. \\
 From $T(A + A) = TA + TA \neq TA$ we obtain that $A + A$ is not isomorphic to $A$.
%$ ~~ $ 2. This property holds for coproducts because $~\perp^0$ is an initial object in $DB$.\\
%3. From 1 and 2.
\\$\square$\\
Notice that for coproducts holds  that $~~ C + \perp^0 ~=~  \perp^0
+ C ~\simeq ~C$, and for any arrow $f$ in $\textbf{DB}$,
  $ ~~f + \perp^1 ~\approx~ \perp^1 + f ~\approx~
  f$, where $~\perp^1~$ is a banal empty morphism between objects, such that
  $~\partial_0(\perp^1)= \partial_1(\perp^1) = \perp^0$, with $\widetilde{\perp^1} =
  \perp^0$.\\
We are ready now to introduce the duality property between
coproducts and products in this $\textbf{DB}$ category:
\begin{propo}\label{prop:co-products} There exists an idempotent coproduct bifunctor
$+:\textbf{DB}\times \textbf{DB}\longrightarrow \textbf{DB}$ which
is a disjoint union
operator for objects and arrows in $\textbf{DB}$.\\
The category $\textbf{DB}$ is cocartesian with initial (zero) object
$~\perp^0$ and for every pair of objects A,B it has a categorial
coproduct $A+B$ with monomorphisms (injections) $in_A:A
\hookrightarrow A+B~$ and $in_B:B \hookrightarrow A+B$.\\
By duality property we have that $\textbf{DB}$ is also cartesian
category with a zero object $~\perp^0$. For each pair of objects
$A,B$ there exists a categorial product $A\times B$ with
epimorphisms (projections) $p_A = in^{OP}_A:A\times A
\twoheadrightarrow A~$ and $p_B = in^{OP}_B:B\times B
\twoheadrightarrow B$, where the product bifunctor is equal to the
coproduct bifunctor, i.e., $ \times~ \equiv~+$.
\end{propo}
%\begin{proof}
\textbf{Proof:} 1. For any identity arrow $(id_A,id_B)$ in
$\textbf{DB}\times \textbf{DB}$, where $id_A ,~ id_b~$ are the
identity arrows of $A$ and $B$ respectively, holds that
$\widetilde{id_A+id_B} = \widetilde{id_A} + \widetilde{id_B} = TA
+TB =T(A+B) = \widetilde{id_{A+B}}$. Thus, $
+^1(id_A,id_B) = id_A+id_B = id_{A+B}$, is an identity arrow of the object $A+B$.\\
2. For any given $k:A\longrightarrow A_1$, $~k_1:A_1\longrightarrow
A_2$, $~l:B\longrightarrow B_1$, $~l_1:B_1\longrightarrow B_2$,
holds $\widetilde{+^1(k_1,l_1)\circ +^1(k,l)} =
\widetilde{+^1(k_1,l_1)}\bigcap \widetilde{+^1(k,l)}= \widetilde{k_1
\circ k + l_1 \circ
 l} = \widetilde{+^1(k_1 \circ k ,~ l_1 \circ l)}\\ = \widetilde{+^1((k_1, k) \circ(l_1,
 l))}$, thus $~+^1(k_1,l_1)\circ +^1(k,l) = +^1((k_1, k) \circ(l_1,
 l)).$\\
3. Let us demonstrate the coproduct property of this bifunctor: for
any two arrows $f:A\longrightarrow C$, $g:B\longrightarrow C$, there
exists a unique arrow $k:A+B\longrightarrow C$, such that $f =
k\circ in_A$, $g = k\circ in_B$, where $in_A:A \hookrightarrow A+B$,
$in_B:B \hookrightarrow A+B~$ are the injection (point to point)
monomorphisms ($\widetilde{in_A} = TA,~\widetilde{in_B}
= TB$). \\
It is easy to verify that for any two arrows $f:A\longrightarrow C$,
$g:B\longrightarrow C$,  there is exactly one arrow  $k = e_C
\circ(f+g):A+B \longrightarrow C$, where $e_C:C+C \twoheadrightarrow
C$ is an epimorphism (with $\widetilde{e_C} = TC$), such that
$\widetilde{k} = \widetilde{f} + \widetilde{g}$.
 %\end{proof}
\\$\square$
%\\
%The opposite operation to (co)product (a DBMS's separation) is the
%DBMS's  \emph{Data federation} of two database instances $A$ and $B$
%is their disjoint union, that is a database $A \biguplus B$.
%Between data federation and coproduct (separation) there is the following relationship:
%\begin{propo} \label{prop:Federat}
%For any three database instances $A,B,C$ in $DB$ it holds that:\\
%$A \biguplus (B +C) \simeq (A \biguplus B) + (A \biguplus C)$.
%\end{propo}
%\textbf{Proof}: it is easy to verify that every view that can be
%obtained from the object $A \biguplus (B +C)$ that is the database
%obtained by federating database $A$ with two separated databases $B$
%and $C$ so that it is impossible to write a query with relations in
%both databases $B$ and $C$,  can be obtained as well from the
%database $(A \biguplus B) + (A \biguplus C)$ that is a separation of
%two federated databases $A \biguplus B$ and $A \biguplus C$. And
%viceversa.\\
%Thus we have that $T(A \biguplus (B +C)) = T((A \biguplus B) + (A
%\biguplus C))$, and, consequently, $A \biguplus (B +C) \simeq (A
%\biguplus B) + (A \biguplus C)$.
%\\$\square$

%\begin{figure}
%\begin{center}
% \includegraphics{fig6.eps}
% \caption{pullback diagram}
% \end{center}
% \end{figure}
%
\subsection{Data Federation operator in DB}
The opposite operation to (co)product (a DBMS's separation) is the
DBMS's  \emph{Data federation} of two database instances $A$ and
$B$.\\
% is their disjoint union, that is a database $A \biguplus B$.
 A federated database system is a type of meta-database management
system (DBMS) which transparently integrates multiple autonomous
database systems into a \emph{single federated database}. The
constituent databases are interconnected via a computer network, and
may be geographically decentralized. Since the constituent database
systems remain autonomous, a federated database system is a
contrastable alternative to the (sometimes daunting) task of merging
together several disparate databases. A federated database, or
virtual database, is the fully-integrated, logical composite of all
constituent databases in a federated database system.\\
In this way we are able to compute the queries with the relations of
\emph{both} databases. In fact, Data Federation technology is just
used for such an integration of two previously separated
databases.\\
Consequently, given any two databases (objects in $\textbf{DB}$) $A$
and $B$, the federation of them (under the common DBMS) corresponds
to \emph{disjoint union} of them under the same DBMS, thus, equal to
database $A \biguplus B$.
%In what follows we will see that data federation of two databases
%(objects in $\textbf{DB}$) is isomorphic to the object in
%$\textbf{DB}$ obtained by the merging of these two original
%database, defined in Section \ref{section:merging}.

%
\section{Categorial semantics of database schema mappings \label{Section:interpret}}
 It is natural for the database schema $\A = (S_A,\Sigma_A )$, where $S_A$ is a set of n-ary relation symbols  and $\Sigma_A$ are the
database integrity constraints, to take $\Sigma_A$ to be a
\emph{tuple-generating dependency (tgd)} and
\emph{equality-generating dependency} (egd). We denote by
$\A_\varnothing$ the empty database schema with empty set of
relation symbols, where $\Sigma_{\A_\varnothing}$
is the empty set of integrity constraints.\\
 A tgd says that if
some tuples, satisfying certain equalities exist in the relation,
then some other tuples (possibly with some unknown values), must
also exist in the relation.\\
An egd says that if some tuples, satisfying certain equalities exist
in the relation, then some values in these tuples must be equal.
Functional dependencies are egd's of a special form, as for example
primary-key integrity constraints.\\
 These two classes of dependencies
together comprise the \emph{embedded implication dependencies} (EID)
~\cite{Fagi82} which seem to include essentially all of the
naturally-occuring constraints on relational databases (the bolded
variables $\textbf{x}, \textbf{y}$ denotes a nonempty list of
variables):
\begin{enumerate}
  \item  a \emph{tuple-generating dependency (tgd)} of the FOL form\\
     $~~~~~~ \forall {\bf x} ~ (\exists {\bf y}~\phi_A(\bf x,\bf y)~ \Rightarrow ~
      \exists {\bf z} ~ (\psi_A(\bf x,\bf z))$\\
      where the formulae $\phi_A(\textbf{x}, \textbf{y})$  and $\psi_A(\textbf{x},\textbf{z})$ are conjunctions of
     atomic formulas over $\A$
     (for integrity constraints over database schemas we will consider only class of weakly-full tgd for which
     query answering is decidable, i.e., when the right-hand side has no existentially
      quantified variables, and if each $y_i \in \bf y$ appears at most once in the left
      side).
     \item  an \emph{equality-generating dependency} (egd):
     $ ~~~~~~\forall {\bf x} ~(\phi_A(\bf x)~ \Rightarrow ~( x_1 = x_2 )) $\\
     where a formula $\phi_A(\textbf{x})$ is a conjunction of
     atomic formulas over $\A$, and $\textbf{x}_1, \textbf{x}_2~$ are among the
     variables in \textbf{x}.
\end{enumerate}
Notice that any schema database \emph{mapping} from a schema $\A$
into a schema $\B$ is represented by the general tgd $~ \forall
{\textbf{x}} ~ (\exists {\textbf{y}}~\phi_A(\textbf{x},\textbf{y})~
\Rightarrow ~
      \exists {\textbf{z}} ~ (\psi_B(\textbf{x},\textbf{z}))$, that
      is by the \emph{view mapping} $~ q_A(\textbf{x})\Rightarrow
      ~q_B(\textbf{x})$, as will be used in what follows in
      Definition \ref{def:graphmaps}, where $q_A(\textbf{x})$ (equivalent to $(\exists {\textbf{y}}~\phi_A(\textbf{x},\textbf{y})$),  is a query over
      the schema $\A$, and $q_B(\textbf{x})$ (equivalent to $(\exists {\textbf{z}}~\psi_B(\textbf{x},\textbf{y})$),  is a query over
      the schema $\B$.\\
 In what follows we will explain how the logical
\emph{model} theory for database schemas and their mappings based on
views, can be translated into the category theory by using the
$\textbf{DB}$ category defined in the previous chapter.
 The integrity constraints for databases are expressed by  the FOL
logical sentences (the FOL formulae without free variables), and
such sentences are expressed in the schema database level by the
mappings from the database schema $\A$ into the empty database
schema $\A_\varnothing$.
%The Yes/No queries are expressed in this way as well.
We define their denotation in the $\textbf{DB}$ category as follows:
\begin{definition} \label{def:sentences1}
 For any sentence  $\varphi:\A \rightarrow \A_\varnothing$ (a logic formula without
variables, in Definition \ref{def:view-map}) over a database schema
$\A$ and a given interpretation $\alpha$ such that $\varphi$ is
satisfied by it, then there exists the unique morphism from $\A$
into terminal object $\perp^0$ in $\textbf{DB}$ category, $f:A
\rightarrow \perp^0$, where $f = \alpha^*(\varphi)$ and $A =
\alpha^*(\A)$ (for $A \simeq \perp^0$ as well). Otherwise, when $A =
\alpha^*(\A)$ is not isomorphic to $\perp^0$, if $\varphi$ is not
satisfied by $\alpha$
 then $\alpha^*(\varphi)$ is mapped into the identity arrow
$~id_{\perp^0}:\perp^0 \rightarrow \perp^0$.
% the following property is
%satisfied: $~~~\widetilde{f} = \perp^0~$ if $~\varphi$ is true for
%this interpretation; empty set, otherwise.
\end{definition}
Notice that, differently from view-mappings for queries (formulae
with free variables) given in Definition \ref{def:atomicmorphisms},
the integrity constraints have in a $\textbf{DB}$ category the empty
database (zero object, i.e., terminal an initial) as the codomain,
and the information flux equal to $\perp^0 = \{\perp\}$. It is
consistent with definition of morphisms in $\textbf{DB}$ category,
because the sentences do not transfer any data from source to target
database, so, their information flux has to be empty. In the case of
ordinary query mappings, the minimal information flux is $\{\perp\}
= \perp^0$ as well. In $\textbf{DB}$ category, for any unique
morphism  from initial object $\perp^0$ (empty database) to another
object (database) $A$, $f:\perp^0 \rightarrow A$, the information
flux of these morphisms is also equal to $\perp^0$.\\
 Based on this semantics for logic formulae
without free variables (integrity constraints and Yes/No queries),
we are able to define the categorial interpretations for database
schema mappings, as follows.
\subsection{Categorial semantics of database schemas}
As we explained in  Section \ref{section:prelim}, in order to define
the database mapping systems we will use two fundamental operators
for the database schemas, data federation $\bigoplus$ and data
separation $\dagger$, with the two correspondent monoids,
$((\mathbb{S}, =), \dagger ,\A_\emptyset)$ and $((\mathbb{S}, =),
\bigoplus,\A_\emptyset)$, and with the distribution low: $\A
\bigoplus(\B \dagger
\C) = (\A \bigoplus \B)\dagger (\A \bigoplus \C)$.\\
Consequently, each vertex in a graph $G$ of a database mapping
system, is a term of the combined algebra of these two monoids,
$\mathbb{S}_{Alg} = ((\mathbb{S}, =), \bigoplus, \dagger,\A_\emptyset)$.\\
In what follows, we say the database schema for any well formed
\emph{term} (i.e., an algebraic expression) of this algebra for
schemas $\mathbb{S}_{Alg}$, and we denote by $\A \in
\mathbb{S}_{Alg}$ a database schema that can be an \emph{atomic}
schema, or composed schema by a finite number of atomic schemas and
two algebraic operators $\bigoplus$ and data separation
$\dagger$ of the algebra $\mathbb{S}_{Alg}$.\\
Consequently for \emph{each} schema $\A \in \mathbb{S}_{Alg}$, we
have that $\A =(S_A,\Sigma_A )$, where $S_A = \biguplus \{S_{B_i}
~|~ \B_i$ is an atomic schema in the schema expression $\A \}$ and
$\Sigma_A = \biguplus \{\Sigma_{B_i} ~|~ \B_i$ is an atomic schema
in the schema expression $\A \}$.\\
For each atomic schema database and an interpretation $\alpha$, we
have that $A = \alpha^*(\A)$ is an instance-database of this schema,
thus, it is an object in $\textbf{DB}$ category. For the composite
schemas (the non atomic terms of the algebra $\mathbb{S}_{Alg}$
their interpretation in $\textbf{DB}$ category is given by the
following Proposition:
\begin{propo} \label{prop:Homomorph} For a given interpretation
$\alpha$ the following homomorphism from schema database level and instance database level there exists:\\
$\alpha^*:((\mathbb{S}, =), \bigoplus, \dagger,\A_\emptyset)
\rightarrow ((Ob_{DB}, \simeq), \biguplus, +,  \perp^0)$.
\end{propo}
\textbf{Proof}: The interpretation of a given schema $\A$ is an
instance $A = \alpha^*(\A)$ of this database, that is an object in
$\textbf{DB}$, while for every interpretation $\alpha^*(\A_\emptyset) = \perp^0$.\\
From the monoidal property we have the equation $\A \bigoplus
\A_\emptyset = \A$ in the algebra $\mathbb{S}_{Alg}$. By the
homomorphism above we have that $\alpha^*(=)~ = ~\simeq,
\alpha^*(\bigoplus) = \biguplus$, so that $\alpha^*(\A \bigoplus
\A_\emptyset) = \alpha^*(\A) \biguplus \alpha^*(\A_\emptyset) = A
\biguplus \perp^0 \simeq A$.
%, and from Proposition \ref{prop-merging} for merging of objects in $DB$ it
%holds the isomorphism $A \oplus \perp^0 \simeq A$.\\
From the monoidal property we have the equation $\A \dagger
\A_\emptyset = \A$ in the algebra $\mathbb{S}_{Alg}$. By the
homomorphism above we have that $\alpha^*(\dagger) = +$, so that
$\alpha^*(\A \dagger \A_\emptyset) = \alpha^*(\A) +
\alpha^*(\A_\emptyset) = A + \perp^0$, and  for coproducts  in
$\textbf{DB}$ it holds the isomorphism $A + \perp^0 \simeq A$.
\\$\square$\\
 Let $\A =(S_A,\Sigma_A )$ be
the database schema,
% expressed in a language $\L_D$ over an alphabet $\A_D$,
where  $S_A$ is a set of relation symbols with a given list of
attributes and $\Sigma_A = \Sigma_A^{tgd} \bigcup \Sigma_A^{egd} =
\pi_2(\A)$ are the
database integrity constraints (set of EIDs) which can be empty st as well.\\
We can represent it by a sketch schema mapping $\phi_A:\A
\longrightarrow \A_\varnothing$ ($\phi_A$  denotes the sentence
obtained by conjunction of
  all formulae in $\Sigma_A$), where $\A_\varnothing$ is the empty schema, such
that for any interpretation $\alpha$ it holds that
$\alpha(\A_\varnothing) = \perp^0$.
%Then, its denotation in
%$\textbf{DB}$ can be given by an arrow as follows:
%
\begin{propo} \label{prop:model} If for a database schema $\A = (S_A,\Sigma_A )$ there
exists a model
%(non empty database instance)
$A$ which satisfies all integrity constraints $\Sigma_A =
\Sigma_A^{tgd} \bigcup \Sigma_A^{egd}$ ($\phi_A$ will denote the
sentence obtained by conjunction of
  all formulae in $\Sigma_A$), then there exists the
following interpretation R-algebra $\alpha$ and its extension, the
functor $\alpha^*:\textbf{Sch}(G)\longrightarrow \textbf{DB}$, where
$\textbf{Sch}(G)$ is the sketch category derived from the graph $G$
with the arrow $\Sigma_A:\A \longrightarrow \A_\varnothing$ (i.e.,
$\textbf{Sch}(G)$ is composed by the objects $\A$, $\A_\varnothing$,
the arrow $\phi_A:\A \longrightarrow \A_\varnothing$ and  the
identity arrows $id_{\A}:\A \longrightarrow \A$ and
$id_{\A_\varnothing}:\A_\varnothing \longrightarrow
\A_\varnothing$), such that:
\begin{enumerate}
  \item $\alpha^*(\A) \triangleq A$, where $A$ is possibly empty database instance (with all empty
  relations) as well
  \item $\alpha^*(id_{\A}) \triangleq id_A:A\longrightarrow A$
  \item $\alpha^*(id_{\A_\varnothing}) \triangleq id_{\perp^0}:\perp^0\longrightarrow \perp^0$
  \item $\alpha^*(\phi_A) \triangleq(f_{tgd} \bigcup f_{egd}):A \longrightarrow
  \perp^0$.
 % , where for each integrity constraint $q_i \in \Sigma_A$,  $\widetilde{\alpha(q_i)} = \perp^0$.
\end{enumerate}
\end{propo}
\textbf{Proof}: from Definition \ref{def:sentences1} and the point 4
of this proposition we have that each integrity constraint $q_i \in
\Sigma_A$ of the database schema $\A$ is satisfied by the
interpretation $\alpha$ (because the conjunction of all integrity
constraints, denoted by $\phi_A$, is satisfied w.r.t the Definition
\ref{def:sentences1}): if $\Sigma_A$ is empty, then it is always
satisfied as usual. Thus $\alpha$ is a model of a database schema
$\A$, and the instance of this model is the nonempty database $A
=\alpha^*(\A)$, that is an object in the $\textbf{DB}$ category.
\\$\square$\\
Notice that any empty database $A$ (such that all its relations are
empty) is isomorphic to the database $\perp^0$ with only one empty
relation $\perp$ (i.e., $\perp^0 = \{\perp\}$). It is easy to show,
based on the fact that any arrow for this empty database $f:A
\rightarrow A$ has the information flux $\widetilde{f} = \perp^0$,
so that $f = id_A$ is the unique identity arrow for this empty
database. But the unique arrows $g:A \rightarrow \perp^0$ and
$h:\perp^0 \rightarrow A$ have the same information fluxes, i.e,
$\widetilde{g} = \widetilde{h} = \perp^0$, so that $g \circ h =
id_{\perp^0}$ and $h \circ g = id_A$,
and, consequently,  $A \simeq \perp^0$.\\
%Consequently, the remark in the point 1 of this Proposition
%specifies that $A = \alpha^*(\A)$ is a non empty database instance
%of the non empty database schema $\A$.
Consequently, the remark in the point 1 of this Proposition
specifies that if $A = \alpha^*(\A) \simeq \perp^0$ is empty
database, than $\alpha^*$ is a model of a schema $\A$, and integrity
constraint for point 4 corresponds to the satisfaction of this
integrity constraint w.r.t. the Definition \ref{def:sentences1}.
\subsection{Categorial semantics of  database mappings}
First of all, we will define formally a schema database mapping
\emph{graph} $G$, as follows:
\begin{definition} \label{def:mapsystem}
A schema database mapping \emph{graph} $G$ is composed by an atomic
arrow $\Sigma_A:\A \rightarrow \A_\emptyset$, for each database
schema $\A$, and a number of (a view based) atomic schema database
mappings $~\M:\A
\rightarrow \B$ between two given schemas $A$ and $\B$.\\
we will use the following basic binary operators for these database
mapping graphs:
\begin{itemize}
  \item
 Given two mappings $~\M_1:\A \rightarrow \B$ and
$~\M_2:\B \rightarrow \C$, we will denote their sequential
composition in this graph $G$ by $~\M_2 ; \M_1$, where $;$ is a
binary associative but non commutative operator.
\item Given two mappings $~\M_1:\A \rightarrow \B$ and $~\M_2:\A
\rightarrow \C$, we will denote their branching in this graph $G$ by
$~\M_2 \biguplus \M_1:\A \rightarrow B \dag C$, where $\biguplus$ is
a binary associative and commutative operator of disjoint union.
\end{itemize}
\end{definition}
This is easy to verify that a graph $G$ can be extended into a
 sketch category $\textbf{Sch}(G)$.\\
   The semantics of  a view-based mapping $~\M = \{q_{Ai}(\textbf{x}_i) \Rightarrow q_{Bi}(\textbf{x}_i) | 1 \leq i \leq n \}$
   from a relational database schema $\A$ into
  a database schema $\B$,
  % $f_{Sch}:\A\longrightarrow \B$,
   is a
  constraint on the pairs of interpretations, of $\A$ and $\B$,
  and therefore specifies which pairs of interpretations can
  co-exist, given the mapping (see also ~\cite{MBDH02}). The
  formalization of the embedding $\gamma:G \rightarrow \textbf{Sch}(G)$ of a graph $G$ into the sketch $\textbf{Sch}(G)$  can be given
  by iteration of the following rules:
  \begin{definition} \label{def:graphmaps}
  We consider the view-based mappings between schemas defined in the SQL language of $SPJRU$
  algebra.
  The arrows in the sketch $\textbf{Sch}(G)$, for any arrow $~\M:\A \rightarrow
\B$ in a given graph $G$ in Definition \ref{def:mapsystem},
  where  $~\M = \{q_{Ai}(\textbf{x}_i) \Rightarrow q_{Bi}(\textbf{x}_i) | 1 \leq i \leq n \}$
  and
  $q_{Ai}(\textbf{x}_i), q_{Bi}(\textbf{x}_i)$ are open FOL formulae  over $\A$, are defined as follows:
\begin{enumerate}
  \item    for each  $q_{Ai}(\textbf{x}_i) \Rightarrow
  q_{Bi}(\textbf{x}_i) \in \M$ such that that $q_{Bi}$ is not a relation symbol  of a database
  schema $\B$, we introduce a new relation $r_i(\textbf{x}_i)$ in $\gamma(\B)$ (we will use the same
  symbol
   for this $\gamma$-enlarged database schema by these new relations).
   Then we introduce in $\textbf{Sch}(G)$ the single mapping arrow $f_{\M}:\A
  \rightarrow \B$, where,
  %\\ $~f_{\M} = \{q_{Ai}(\textbf{x}_i) \Rightarrow r_i(\textbf{x}_i) | 1 \leq i \leq n $ and $r_i$ is a relation symbol in $\B
  %\}$,  i.e., an atomic schema morphism
\[
   f_{\M} = \gamma(\M) \triangleq \bigcup_{1 \leq i \leq n } \{v_i\cdot q_{i}~: \A\longrightarrow
   \B ~|~  \partial_0(q_{i})= R_{i}
   ,~ \partial_1(q_{i})= \partial_0(v_i)
   ,~ \partial_1(v_i) = \{ r_i\} \}
   \]
where $q_i, v_i$ are abstract "operations" (operads) introduced in
Definition \ref{def:atomicmorphisms}, such that for a given model
$\alpha$ of this database schema mapping, $\alpha(q_i)$ is a query
computation of a query $q_{Ai}(\textbf{x}_i)$. The set $R_{i}$ is
the set of relation symbols in $\A$ used in the formula
$q_{Ai}(\textbf{x}_i)$.
%
%   in the way that all second components of query
%  implications in $g$ are simple relations and not complex queries.
%
  \item  for each $q_{Ai}(\textbf{x}_i) \Rightarrow
  q_{Bi}(\textbf{x}_i) \in \M$ such that
  $q_{Bi}$  is not a relation symbol  of a database
  schema $\B$ (then $ ~q_{Ai}(\textbf{x}_i) \Rightarrow
  q_{Bi}(\textbf{x}_i)$ (logical implication between queries) means that each tuple of the view obtained by the
  query $q_{Ai}(\textbf{x}_i)$ is also a tuple of the view obtained by the
  query $q_{Bi}(\textbf{x}_i)$), we do as follows:\\
   We introduce in this sketch $\textbf{Sch}(G)$
  a new helper database schema $\C_i$ with a single relation $c_i(\textbf{x}_i,y)$, and   two new
  schema mappings:\\ $f_{AC_i} = w_i\cdot q_{i}:\A\rightarrow \C_i$ (with $\partial_0(w_i) =
  \partial_1(q_{i})
   ,~ \partial_1(w_i) = \{ c_i\}$ ), and\\ $f_{BC_i} = w'_i\cdot q'_{i}:\B\rightarrow \C_i$ (with $\partial_0(q'_{i})= R'_{i}
   ,~ \partial_1(q'_{i})= \partial_0(w'_i)
   ,~ \partial_1(w'_i) = \{ c_i\}$, where $R'_{i}$ is the set of all relation  symbols in $\B$ used in the formula
$q_{Bi}(\textbf{x}_i)$),
  such that
   $f_{AC_i}$ corresponds to $\{q_{Ai}(\textbf{x}_i) \Rightarrow c_i(\textbf{x}_i,\natural_A)\}$ and
    $f_{BC_i}$ corresponds to $\{q_{Bi}(\textbf{x}_i) \Rightarrow c_i(\textbf{x}_i,\natural_B)\}$,
   where $\natural_A, \natural_B$ are two new values not present in the dominium of the
   databases. Consequently, we introduce in $G$ also the integrity
   constraint arrow
   $\varphi_i:\C_i \rightarrow \A_\varnothing$ for this new schema $\C_i$,
   where the sentence $\varphi_i$ is equal to the tgd $~ \forall {\textbf{x}_i} ~
   (\exists y (c_i(\textbf{x}_i, y) \wedge y = \natural_A)~ \Longrightarrow ~
      \exists z(c_i(\textbf{x}_i, z)\wedge z = \natural_B))$.
  \end{enumerate}
 \end{definition}
% Notice that this general definition of the schema mapping graph $G$
% can be simplified in certain cases, as shown in Section \ref{Section:GLAV} for GLAV data integration, but by imposing additional
% properties for the functorial translation into DB category (in Theorem  \ref{th:GLAV}).\\
 It is easy to verify that in the obtained sketch $\textbf{Sch}(G)$, between given any two
 nodes there is at maximum \emph{one} arrow.
 There is a fundamental functorial \emph{interpretation} connection from schema
 mappings and their models in the instance level category $\textbf{DB}$:
 based on the Lawvere categorial theories \cite{Lawve63,BaWe85}, where he
 introduced a way of describing algebraic structures using
 categories for theories, functors (into base category $\textbf{Set}$, which
 we will substitute by more adequate category $\textbf{DB}$), and natural
 transformations for morphisms between models.\\ For example, Lawvere's seminal
 observation that the theory of groups is a category
 with group object, that group in $\textbf{Set}$ is a product preserving
 functor, and that a morphism of groups is a natural transformation
 of functors, is an original new idea that was successively
 extended in order to define the categorial semantics for different
 algebraic and logic theories.\\ This work is based on the theory of
 \emph{sketches}, which are fundamentally small categories obtained from graphs enriched by
 concepts such as (co)cones mapped by functors in (co)limits of the base
 category $\textbf{Set}$. It was demonstrated that, for every sentence in
 basic logic, there is a sketch with the same category of models, and
 vice versa \cite{Mapa89}. Accordingly, sketches are called
 graph-based logic and provide very clear and intuitive
 specification of computational data and activities. For any small sketch
 $\textbf{E}$ the category of models $Mod(\textbf{E})$ is an accessible category by Lair's theorem and
 reflexive subcategory of $\textbf{Set}^{\textbf{E}}$ by Ehresmann-Kennison theorem.
A generalization to  base categories other than $\textbf{Set}$ was
proved by Freyd and Kelly (1972).
 In
 what follows we will substitute the base category $\textbf{Set}$ by this new
 database category $\textbf{DB}$.\\
 For instance, for the separation-composition mapping cocone diagram (graph $G$), given
 in the introduction, its translation in a \emph{sketch} (a category $\textbf{Sch}(G)$) is presented in the left commutative diagram below
 (notice that mapping arrow $\M$ in a graph $G$ ar replaced by the morphism $f_{\M}$ in this sketch, while the nodes (objects) are changed
  eventually by introducing another auxiliary relation symbols as explained in Definition \ref{def:graphmaps}),
 and the functorial translation of this sketch into $\textbf{DB}$ category
 has to be coproduct diagram in $\textbf{DB}$
 %(in Definition \ref{def:coproduct} and  Proposition \ref{prop:co-products})
  as follows:
\begin{diagram}
 \gamma(\A)        & \rTo{f_{\M_1}}          &  \gamma(\A) \dagger \gamma(\B)     &  \lTo{f_{\M_2}}    & \gamma(\B)           &  &A        & \rInto{In_A}          &  A + B     &  \lInto{In_B}    & B \\
           & \rdTo_{f_{\M_A}}        &   \dTo{f_{\M}}               &  \ldTo_{f_{\M_B}}  &  & \Rrightarrow  &        & \rdTo_{f}        &   \dTo{k}               &  \ldTo_{g}  &\\
           &                     &   \gamma(\C)             &              &           & &  &                   &   C             &                &\\
\end{diagram}
As we explained in the introduction, in database mapping systems,
expressed by a graph $G$, we will never use "commutative diagrams"
as left diagram above (but only an arrow $f_{\M}:\gamma(\A )\dagger
\gamma(\B) \rightarrow \gamma(\C)$, or, more frequently, two simple
arrows $f_{\M_A} = \gamma(\M_A):\gamma(\A) \rightarrow \gamma(\C)$
and $f_{\M_B} = \gamma(\M_B):\gamma(\B) \rightarrow \gamma(\C)$),
our sketch $\textbf{E} = \textbf{Sch}(G)$ will be a simple small
category, i.e., 4-tuple $(G,u,D,C)$ where $D$ and $C$ are empty
sets. Consequently, these database-mapping sketches are more simple
than the sketches used for definition of Entity-Relationship models
of single relational databases.
 \begin{propo} \label{prop:schema}
 Let $\textbf{Sch}(G)~$ be a schema sketch category generated from a schema
 mapping graph  $G$, obtained by applying method in Definition \ref{def:graphmaps} for each mapping between two database schemas in a given
 database mapping system with $n \geq 2$ database schemas. Let an interpretation R-algebra
 $\alpha~$ satisfies the following property:
 for any database schema $\A $, (object in
 $\textbf{Sch}(G)$), $\alpha$ satisfies the Proposition \ref{prop:model}, so that $A \triangleq \alpha^*(\A) \in Ob_{DB}$ is a  model of the database schema
  $\A$, and for each schema mapping arrow $f_{Sch}:\A\longrightarrow \B$, (where $\B$ is not empty schema) the
 atomic morphism in $\textbf{DB}$ category $\alpha^*(f_{Sch}):\alpha^*(\A) \rightarrow \alpha^*(\B)$ is determined
 by banal  \textsl{set-inclusion case} of Definition \ref{def:atomicmorphisms}.\\
  Then there is the functor (categorial model) $~\alpha^*:\textbf{Sch}(G)\longrightarrow
 \textbf{DB}~$.
 The set of categorial models of the database schema mapping graph
 $G$ is equal to the homset $hom(\textbf{Sch}(G), \textbf{DB})$ of all functors from these two categories in the category $~\textbf{Cat}$,
  i.e. equal to the set of all objects in the category of functors $~\textbf{DB}^{\textbf{Sch}(G)}$ as well.\\
 For a given model (functor) $~\alpha^* \in \textbf{DB}^{\textbf{Sch}(G)}$, its image in $\textbf{DB}$  will be called
 a \verb"DB-mapping"\\ \verb"system", and denoted by $\M_S$.
  \end{propo}
% \begin{proof}
 \textbf{Proof:}
 This is easy to verify, based on general theory for sketches \cite{BaWe85}: each arrow in a sketch
 (obtained from a schema mapping graph $G$) may be
 converted into a tree syntax structure of some morphism in $\textbf{DB}$ (labeled tree without
 any interpretation).
 %, thus, a graph $G$ can be extended into a
 %sketch category $\textbf{Sch}(G)$.
 %%with the analog composition of arrows as in $DB$
 %(The composition of schema mappings in the category $\textbf{Sch}(G)$,
 %where each mapping is a set of first-order logical formulas,
 %can be defined as a disjoint union).
 The functor $\alpha^*$ is
 only the simple extension of the interpretation R-algebra function
 $\alpha~$ for a lists of symbols. The functorial property for the identity mappings follows from Proposition  \ref{prop:model} and for two
 atomic mappings $f_{Sch}:\A\longrightarrow \B$, $g_{Sch}:\B\longrightarrow
 \C$, and their atomic morphisms in $\textbf{DB}$, $f = \alpha^*(f_{Sch})$, $g = \alpha^*(g_{Sch})$, we have that $\alpha^*(g_{Sch} \circ f_{Sch}) =
 g \circ f$ as defined in Definition \ref{def:morphisms}.
 It remains only to verify that for each auxiliary database schema
 $\C_i$ and integrity constraint $\varphi_i:\C_i \rightarrow
 \A_\emptyset$ (in Definition \ref{def:graphmaps}) the operator
 $\alpha^*$ satisfies the functorial property, such that this schema
 arrow is mapped into the arrow $\alpha^*(\varphi_i):C_i \rightarrow
 \perp^0$, where $C_i = \alpha^*(\C_i)$. In fact it holds, because
 if $\alpha$ is a model of this database mapping system represented
 by the graph $G$, then this integrity constraint $\varphi_i$ is
 satisfied, and based on the Definition \ref{def:sentences1} the
 functorial property is satisfied.\\ Notice that this is true also
 in special cases when $ C_i = \alpha^*(\C_i) ~\simeq
 ~\perp^0$. This case happens when for a view mapping $q_A(\textbf{x}) \Rightarrow
 q_B(\textbf{x})$, both $\|q_A(\textbf{x})\|$ (resulting view of the query $q_A(\textbf{x})$ over the database $A = \alpha^*(\A)$),
  and $\|q_B(\textbf{x})\|$ (resulting view of the query $q_B(\textbf{x})$ over the database
  $B = \alpha^*(\B)$), are empty relations, and consequently $\alpha(c_i(\textbf{x},y))$ is empty
  relation in the database instance $C_i = \alpha^*(\C_i) = \{\alpha(c_i(\textbf{x},y))\}$, so that $C_i \simeq
  \perp^0$.\\
  Thus an integrity constraint $\varphi_i:\C_i \rightarrow \A_\emptyset$ (an auxiliary
  arrow in $\textbf{Sch}(G)$ obtained from some mapping between two database
  schemas in $G$) can be unsatisfied only if $C_i = \alpha^*(\C_i)$
  is not isomorphic to  $\perp^0$.\\
  In order to prove that the set of functors in  $\textbf{DB}^{\textbf{Sch}(G)}$ is
  exactly the set of al models of the database mapping system
  expressed by the graph $G$, it is now enough to prove that any
  interpretation $\alpha$ that \emph{is not} a model of $G$, then
  can not be a functor from $\textbf{Sch}(G)$ into $\textbf{DB}$.
  In order that a given $\alpha$ is not a model of $G$, must be
  satisfied one of the following cases:\\
  1. case when for some database schema $\A$ in $G$, $\alpha^*(\A)$
  is not a model of this database: it means that the conjunction of all integrity
  constraints $\Sigma_A:\A \rightarrow \A_\emptyset$ of $\A$ is not satisfied by $\alpha$ (thus when $\alpha^*(\A)$ not isomorphic to $\perp^0$, as
  specified by Definition \ref{def:sentences1}), so that from Definition \ref{def:sentences1},
  it holds that $\alpha^*(\Sigma_a) = id_{\perp^0}:\perp^0 \rightarrow
  \perp^0$, and it does not satisfy the functorial requirement
  because database instance $\alpha^*(\A)$ is not isomorphic to
  $\perp^0$.\\
  2. case when some  integrity constraint $\varphi_i:\C_i \rightarrow
  \A_\emptyset$ (an auxiliary
  arrow in $\textbf{Sch}(G)$ obtained from some mapping between two database
  schemas in $G$),
  with $C_i = \alpha^*(\C_i)$ is not isomorphic to $\perp^0$, is not
  satisfied by $\alpha$, so that from Definition \ref{def:sentences1} it holds that $\alpha^*(\varphi_i) = id_{\perp^0}:\perp^0 \rightarrow
  \perp^0$, and it does not satisfy the functorial requirement
  (because database instance $\alpha^*(\C_i)$ is not isomorphic to
  $\perp^0$).
% \end{proof}
\\$\square$\\
Notice that in this functorial semantics for database mappings from
an original schema database mapping  $\M:\A \rightarrow \B$, with a
correspondent arrow $f_{\M} = \gamma(\M):\gamma(\A) \rightarrow
\gamma(\B)$, where $\gamma(\A), \gamma(\B)$ are the original
database schemas enlarged by a number of auxiliary relations
introduced in Definition \ref{def:graphmaps}, can be translated into
the arrow $f = \alpha^*(\gamma(\M)):\alpha^*(\A) \rightarrow
\alpha^*(\B)$ between the instances of the original schema databases
$\A$ and $\B$ without added auxiliary relations, because we have
that $\alpha^*(\A), \alpha^*(\A)$ are isomorphic in $\textbf{DB}$ to
$\alpha^*(\gamma(\A)), \alpha^*(\gamma(\A))$, respectively, as
follows:
\begin{propo} \label{prop:isomobj}
For any  database schema $\A$ and $\B$, in a given schema database
mapping graph $G$, it holds that $A = \alpha^*(\A) \simeq
\alpha^*(\gamma(\A))$.\\
Consequently, any functorial semantics of a given schema database
mapping $\M:\A \rightarrow \B$ is represented in the $\textbf{DB}$
category by the morphism, $f \approx
\alpha^*(\gamma(\M)):\alpha^*(\A) \rightarrow \alpha^*(\B)$.
\end{propo}
\textbf{Proof}: It is easy to show that $T(\alpha^*(\A)) =
T(\alpha^*(\gamma(\A)))$, because each $\gamma$-added relation
$r_i(\textbf{x}_i)$ is just a subrelation of the  view obtained by
the query $q_{B_i}(\textbf{x}_i)$ over the relations in the original
database $\B$, that is part of the view mapping
$q_{A_i}(\textbf{x}_i)\Rightarrow q_{B_i}(\textbf{x}_i) \in \M:\A
\rightarrow \B$ (see point 2 in
Definition \ref{def:graphmaps}).\\
Consequently, we have the isomorphisms $is_A: \alpha^*(\A) \simeq
\alpha^*(\gamma(\A))$, $is_B: \alpha^*(\B) \simeq
\alpha^*(\gamma(\B))$, so that, $f = is_B^{-1}\circ
\alpha^*(\gamma(\M))\circ is_A: A \rightarrow B$, i.e., $f \approx
\alpha^*(\gamma(\M))$.
\\$\square$\\
That is the reason that instead of $\gamma(\A)$ we can use the
original database schemas $\A$ and their database instances $A =
\alpha^*(\A)$ in the $\textbf{DB}$ category.
\section{Conclusions}
In previous work we defined a base database category $\textbf{DB}$
where objects are instance-databases and morphisms between them are
extensional GLAV mappings between databases. We defined equivalent
(categorically isomorphic) objects (database instances) from the
\emph{behavioral point of view based on observations}:  each arrow
(morphism) is composed by a number of "queries" (view-maps), and
each query may be seen as an \emph{observation} over some database
instance (object of $\textbf{DB}$). Thus, we  characterized each
object in $\textbf{DB}$ (a database instance) by its behavior
according to a given set of observations. In this way two databases
$A$ and $B$ are equivalent (bisimilar) if they have the same set of
its observable internal states, i.e. when $TA$ is equal to $TB$. It
has been shown that such a $DB$ category is equal to its dual, it is
symmetric in the way that the semantics of each morphism is an
closed object (database) and viceversa each database can be
represented by its identity morphism, so that $\textbf{DB}$ is a
2-category.\\
 In \cite{Majk09a,Majk09a} has been introduced the categorial (functors)
semantics for two basic database operations: \emph{matching} and
\emph{merging} (and data federation), and has been defined the
algebraic database lattice.\\
Here we considered the \emph{schema} level for databases and their
view-based mappings, based on queries. The fundamental operations
for databases in the view of inter-mappings between them is the fact
if they are separated od federated databases. It depends on the kind
of DBMS system used for two mapped databases: when two databases are
federated then we can compute the queries over the relations  of
both databases; when they are separated by two independent DBMS,
then DBMS can compute only the queries with all relations of only
one of these two databases.\\
We have shown that these two fundamental operators, data separation
and data federation, used in schema database mapping system, need a
different base category from $\textbf{Set}$ where coproducts are
equal to products (up to isomorphism). Then we defined the Graphs
schema database mapping systems, and the sketches for such database
graphs. Consequently we defined the categorical functorial semantics
for these sketches into new base database category $\textbf{DB}$.

\bibliographystyle{IEEEbib}

\bibliography{medium-string,krdb,mydb}

%\balancecolumns

\end{document}